\documentclass[acmsmall,screen,nonacm]{acmart}

\usepackage{float}
\usepackage{graphicx}
\usepackage{algorithm}
\usepackage{algpseudocode}
\usepackage{amsmath}
\usepackage{balance}
\usepackage{newtxmath}
\usepackage{listings}
\usepackage{hyperref}
\usepackage{xcolor} 
\usepackage{multirow}
\usepackage{booktabs}
\usepackage{tabularx}
\usepackage{adjustbox}

\AtBeginDocument{%
  }
\setcopyright{acmlicensed}
\copyrightyear{2025}
\acmYear{2025}
\acmDOI{10.1145/3735452.3735528}
\acmConference[PLDI '26]{Proceedings of the 47th ACM SIGPLAN Conference on Programming Language Design and Implementation}{June 15--19, 2026}{Boulder, Colorado, United States}
\acmBooktitle{Proceedings of the 47th ACM SIGPLAN Conference on Programming Language Design and Implementation (PLDI '26), June 15--19, 2026, Boulder, Colorado, United States}
\received{2025-03-21}
\received[accepted]{2025-04-21}
  
\acmISBN{979-8-4007-1921-9/25/06}
\begin{document}
\title{MileStone: A Multi-Objective Compiler Phase Ordering Framework for Graph-based IR-Level Optimization }

\author{Amirhossein Sadr}
\orcid{}
\affiliation{%
  \institution{Faculty of Computer Science and Engineering, \\Shahid Beheshti University}
  \city{Tehran}
  \country{Iran}
}
\email{}

\author{Mehran Alidoost Nia}
\orcid{0000-0002-7274-9569}
\affiliation{%
  \institution{Faculty of Computer Science and Engineering, \\Shahid Beheshti University}
  \city{Tehran}
  \country{Iran}
}
\email{alidoostnia@sbu.ac.ir}

\renewcommand{\shortauthors}{A. Sadr and M. Alidoost Nia}

\begin{abstract}
Compiler phase ordering has a strong effect on program performance. Finding an effective sequence of passes is still a difficult task because the search space is large and execution time, code size and energy consumption often conflict. Existing methods usually depend on fixed optimization levels or limited heuristics and they rarely handle multiple objectives at the same time. This paper presents MileStone, a modular framework that models compiler phase ordering as a multi-objective optimization problem. MileStone represents programs as graphs, predicts performance metrics with a graph neural network and explores pass sequences with a reinforcement-learning agent that follows user constraints. The framework also builds a self-evolving database that collects compiler transformations and improves prediction quality. Experiments on standard benchmarks show that MileStone finds strong Pareto-optimal solutions, meets energy limits more accurately than LLVM optimization levels and other related techniques. MileStone reduces execution time by up to 45 percent under the same energy budget using a multi-objective approach. The results show that MileStone provides an effective and scalable solution for multi-objective compiler phase ordering.
\end{abstract}



\maketitle

\section{Introduction}
Compiler optimization is a fundamental component of modern computing that influences nearly every layer of software performance. It refines high-level code into efficient machine instructions, ensuring programs run faster, consume less energy and use hardware resources more effectively. Whether in mobile devices, cloud platforms or embedded systems, compiler optimizations shape the responsiveness, scalability and energy efficiency of applications. Through techniques such as loop unrolling~\cite{unroll}, inlining~\cite{inline} and vectorization~\cite{vector}, compilers transform programs to achieve better runtime behavior while preserving correctness. 

Modern compiler frameworks are increasingly built around modular architectures that separate analysis, optimization and code generation into reusable components~\cite{modern}. This design philosophy allows developers to assemble, extend or replace individual modules without reengineering the entire compilation pipeline. As a result, researchers and engineers can easily experiment with new optimization techniques, integrate custom analyses or determine compilation strategies to specific hardware or application domains. Frameworks such as LLVM~\cite{llvm,llvm2} exemplify this modularity through their pass-based design and intermediate representation (IR), which enable developers to insert domain-specific transformations or build entirely new compilers by reusing existing infrastructure~\cite{reuse}. Within this modular structure, optimization phases serve as the central mechanism that drives performance enhancement~\cite{opt1}, as their sequence and interaction directly determine how effectively the compiler refines program behavior and exploits hardware capabilities.

The phase ordering problem remains one of the open challenges in compiler optimization research~\cite{phase-prob}. It arises from the need to determine the most effective sequence of optimization passes that transform a program’s IR into efficient machine code. Even though standard optimization levels such as \texttt{-O1}, \texttt{-O2} and \texttt{-O3} offer preconfigured sets of transformations~\cite{phase-o}, they represent only a few points within an enormous and complex search space of possible pass combinations. The order and interaction of these passes influence the final performance, code size and energy consumption of a program, yet there is no universal sequence that consistently leads to optimal results across all applications and architectures~\cite{phase1}. This sort of possibilities makes exhaustive exploration infeasible and highlights why the phase ordering problem continues to be an active area of research in the compiler community.

The core problem addressed in this research lies in the challenge of effectively optimizing compiler pass sequences while simultaneously balancing multiple, often conflicting performance objectives such as execution time, code size and energy consumption. Traditional compilers rely on fixed optimization levels like \texttt{-O3}, which apply predetermined sequences of passes and fail to capture the trade-offs required by diverse programs and architectures~\cite{classic1,classic2}. The complexity of possible phase orderings makes manual heuristic design impractical, while existing learning-based approaches often focus on single-objective optimization or depend on dynamic profiling~\cite{multi1, multi2}. To overcome these limitations, our research formulates compiler phase ordering as a multi-objective optimization problem and explores the use of artificial intelligence (AI) methods~\cite{ai} for performance modeling and adaptive exploration. 

To address these challenges, we introduce MileStone~\footnote{The name MileStone reflects that the framework structures compiler optimization into a sequence of distinct milestones, from graph extraction and database construction to prediction and multi-objective exploration, and identifies Pareto-optimal “milestone” points in the trade-off space of execution time, code size and energy.}, a modular and learning-driven compiler optimization framework designed to explore the phase ordering problem from a multi-objective perspective. MileStone leverages graph-based representations of IR code to capture the structural and semantic relationships within programs and integrates AI components to guide optimization. Specifically, it combines a graph neural network (GNN) for static performance prediction with a reinforcement learning (RL) engine for adaptive exploration of compiler pass sequences. This synergy allows MileStone to efficiently navigate the vast optimization space, discovering Pareto-optimal trade-offs among execution time, code size and energy consumption without requiring exhaustive search or dynamic profiling. The main contributions of the research are as follows:

\begin{itemize}
    \item We introduce MileStone, a unified framework that supports multi-objective compiler phase ordering as an optimization problem balancing execution time, code size and energy consumption.
    \item We present a graph-based neural model that learns structural and semantic properties from IR graphs to accurately predict program performance.
    \item The framework employs an RL-based approach to explore compiler pass sequences and identify Pareto-optimal trade-offs among multiple performance objectives.
    \item The proposed solution incorporates a self-improving data collection mechanism that enhances both predictive accuracy and exploration efficiency.
\end{itemize}

The remainder of paper is organized as follows: In the next section, we review the compiler background needed for this paper. In Section~\ref{sec:problem}, we formally define the main problem of the paper. Section~\ref{sec:system} introduces the MileStone architecture. Section~\ref{sec:exp} reveals the experimental results. Section~\ref{sec:rel} presents the related work that are comparable to the current research. Finally, in Section~\ref{sec:con}, the paper concludes and future directions are presented.

\section{Background}\label{sec:back}
This section reviews the essential background of the proposed MileStone framework including standardized optimization levels, graph-based program representations, learning-based modeling techniques and multiobjective optimization principles. 
\subsection{-O3 Optimization Level}

In a typical compiler architecture, code optimizations are structured as discrete, reusable components known as \textit{passes}. These passes, often derived from common base classes like \texttt{ModulePass} or \texttt{FunctionPass}, are designed to perform specific transformations or analyses on the code~\cite{back1}. This object-oriented approach allows developers to extend the compiler by creating custom passes, which enable fine-grained control over the optimization process.

Compilers generally provide several predefined \textit{optimization levels} such as \texttt{-O1}, \texttt{-O2} and \texttt{-O3}, which represent a subset of sequences from these passes. Each level activates a set of optimizations. The baseline level, often \texttt{-O0}, typically disables most optimizations to ensure fast compilation and ease of debugging~\cite{back2}. In contrast, higher levels introduce increasingly sophisticated transformations aimed at improving execution speed and reducing code size.

The \texttt{-O3} optimization level represents the highest tier of automatic code optimization in most modern compiler toolchains, including GCC, LLVM/Clang and the Intel C++ Compiler. It is designed to maximize runtime performance, often at the cost of increased compilation time and final code size. While lower optimization levels like \texttt{-O1} and \texttt{-O2} focus on a balanced set of safe and efficient transformations, \texttt{-O3} enables an aggressive suite of algorithms. These typically include advanced loop unrolling, function inlining and crucially, auto-vectorization to exploit data-level parallelism using SIMD (Single Instruction, Multiple Data) instruction sets~\cite{back3}. Consequently, \texttt{-O3} is the standard choice for performance-critical release builds where peak execution speed is the primary objective, allowing the compiler to explore a broader and more computationally intensive region of the optimization space.

\subsection{Graph-Based Program Representation}

A well-established approach to program analysis involves representing code as a graph structure, specifically by extracting a Control and Data Flow Graph (CDFG) from a compiler's IR code, such as LLVM IR [8]. This methodology shifts the focus from the syntactic grammar of the source code to capturing the underlying semantics of program execution. In a CDFG, nodes correspond to individual LLVM instructions. The connectivity between these nodes is defined by two fundamental principles: first, edges are drawn to represent the possible paths of control flow, and second, a distinct set of edges is added to represent data dependencies based on the operands of the instructions. This dual representation provides a comprehensive view of program behavior. Furthermore, because the CDFG is derived from the LLVM IR~\cite{back4}, it inherently includes low-level operations (e.g., related to memory management and hardware-oriented instructions).

\subsection{Graph Neural Networks}

Graph data are inherently irregular and non-Euclidean—nodes, and may have varying numbers of neighbors and no consistent spatial ordering. Traditional deep learning models, including convolutional and recurrent neural networks, are therefore ill-suited to directly process such structured graph data~\cite{back5}. To address this limitation, GNNs have been developed as a framework for learning representations directly from graph-structured inputs. The key idea behind GNNs is to enable each node to learn from its local neighborhood through an iterative process known as message passing or neighborhood aggregation. In this process, nodes exchange information with their connected neighbors, update their internal representations and gradually capture both local and global structural dependencies within the graph~\cite{back6}. Through multiple layers of aggregation, GNNs can encode rich contextual information that reflect the topology and feature distribution of the entire graph.

Among the various architectures in this family, \textit{Graph Convolutional Networks} (GCNs) have become the most influential and widely used models~\cite{back7}, which utilize a simple aggregation function that computes a weighted sum of the embeddings of the neighboring nodes with the weights determined by the degree (\(d_i\)) of the nodes, shown by Formula~\ref{eq:GCN}:

\begin{equation}\label{eq:GCN}
h_i' = \sigma \left(W \sum_{j \in N(i) \cup \{i\}} \frac{1}{\sqrt{d_j d_i}} h_j \right)
\end{equation}

\noindent where \(h_i \in \mathbb{R}^F\) (\(h_i' \in \mathbb{R}^{F'}\)) denotes the input (output) embedding of node \(i\), which is a vector of \(F\) (\(F'\)) features. \(W\) is a learnable weight matrix for the transformation function (TF) and \(\sigma\) represents an activation function to introduce non-linearity into the model.

\subsection{Reinforcement Learning} 

Reinforcement Learning or simply RL is a machine learning paradigm in which agents learn optimal decision-making strategies through direct interaction with an environment~\cite{back8}. In RL, the interaction occurs over discrete time steps. At each step $t$, the agent observes the current state $s_t$, selects an action $a_t$, receives a reward $r_t$, and transitions to a new state $s_{t+1}$. The goal of the agent is to maximize the cumulative long-term reward, formulated within the discounted return framework. Unlike supervised learning, which depends on labeled input-output pairs, RL provides feedback only in the form of delayed rewards, making it especially suitable for sequential decision-making under uncertainty~\cite{back9}. Formally, the RL is commonly defined over a Markov Decision Process (MDP), expressed by the tuple $(\mathcal{S}, \mathcal{A}, P, R, \gamma)$, where:
\begin{itemize}
    \item $\mathcal{S}$ denotes the set of possible states,
    \item $\mathcal{A}$ represents the set of available actions,
    \item $P(s'|s, a)$ formulates the transition probability of moving from state $s$ to $s'$ after taking action $a$,
    \item $R(s, a)$ delivers the immediate reward obtained by executing action $a$ in state $s$, and
    \item $\gamma \in [0,1]$ shows the discount factor that controls the importance of future rewards.
\end{itemize}

The fundamental challenge in RL lies in discovering a policy $\pi(a|s)$, which maps states to actions in order to maximize the expected cumulative reward~\cite{back10}. This challenge is typically addressed using either value-based or policy-based approaches. In value-based methods, such as Q-learning~\cite{back11}, the agent learns an action-value function $Q(s,a)$, which estimates the expected return of performing action $a$ in state $s$. The optimal policy is then derived by selecting actions that maximize this value function.

\subsection{Constrained Multiobjective Optimization} 

Constrained multi-objective optimization (CMOO) involves optimizing multiple conflicting objectives simultaneously while satisfying a set of problem-specific constraints~\cite{back12}. Unlike single-objective problems that yield a single optimal solution, CMOO produces a set of trade-off solutions known as the \textit{Pareto-optimal set}, where improving one objective inevitably degrades another~\cite{back13}. Constraints, which may represent physical limits, safety requirements or resource capacities, further restrict the feasible search space and increase problem complexity.

To address such challenges, various approaches have been proposed. Classical scalarization techniques, such as the weighted-sum or $\varepsilon$-constraint methods, transform the multi-objective problem into a single-objective one but often fail to capture diverse Pareto-optimal solutions. In contrast, evolutionary algorithms like NSGA-II, SPEA2 and MOEA/D are widely used for their ability to approximate the entire Pareto front while effectively handling nonlinear and discontinuous constraints through adaptive penalty or repair mechanisms~\cite{back14}. More recently, reinforcement learning and surrogate-assisted optimization have been explored to enhance efficiency in high-dimensional and dynamic environments~\cite{back15}.

\section{Motivation and Problem Definition}\label{sec:problem}

\subsection{Motivation}

The compiler phase ordering problem has traditionally been treated as a single-objective optimization task, often focusing on minimizing either execution time or code size in isolation. However, real-world compilation scenarios rarely prioritize a single metric. Modern applications ranging from high-performance computing to ultra-low-power embedded systems, demand balanced trade-offs among execution time, code size and energy consumption~\cite{motivation}. These objectives are inherently conflicting; for instance, aggressive loop unrolling or vectorization can reduce execution time but simultaneously increase code size and power consumption. As a result, optimizing for one objective may inadvertently degrade others, leading to suboptimal or infeasible outcomes for specific deployment environments. This observation motivates the formulation of the compiler phase ordering problem as a multi-objective optimization problem that explores the Pareto frontier among competing performance metrics rather than a single optimization axis.

Existing works generally collapse objectives into a weighted sum or tune one objective while treating others as secondary constraints. Such formulations fail to capture the Pareto-optimal trade-offs\footnote{The set of non-dominated solutions where improvement in one metric inevitably harms another.}. Modeling the phase ordering problem as a true multi-objective optimization task allows the exploration of this trade-off surface, and enables compilers to adapt optimization sequences to diverse hardware targets and workload requirements. This paradigm opens the way for intelligent compilers that can dynamically choose optimization strategies according to application-specific priorities.

Furthermore, emerging computing platforms, from embedded systems and edge AI devices to data-center servers, demand compilation strategies that respect user-defined constraints. For instance, an embedded system may prioritize energy efficiency over raw execution speed due to strict power budgets. Consider the following example of sequences:

\begin{itemize}
    \item Sequence A: Execution time = 1.2 s, Energy consumption = 5 J.
    \item Sequence B: Execution time = 1.4 s, Energy consumption = 2 J.
\end{itemize}

For a battery-operated microcontroller or IoT node, Sequence B is clearly preferable under a user-specified energy constraint, even though it is slower. Conventional single-objective formulations would fail to recognize such trade-offs. Therefore, a user-constrained multi-objective formulation enables the compiler to adapt optimization decisions to both workload context and platform constraints, and automatically explore the Pareto frontier under user-specified constraint. 

\subsection{Problem Definition}

The main problem of the research is built around the multi-objective phase ordering. To formally shape the problem, we define a multi-objective phase ordering optimization function as follows:
\begin{equation}\label{eq:problem}
U(\text{CodeSize}, \text{ExecTime} \mid \text{Energy}_{\text{target}}) = \mu \frac{\text{CodeSize}}{q} + (1 - \mu)~\text{ExecTime}
\end{equation}
where $0\leq\mu\leq1$ represents the relative importance assigned to different metrics, and $q$ (a finite positive integer) is a scaling constant that balances the numerical ranges of code size and execution time. The objective is to minimize $U$, thereby reducing both code size and execution time while satisfying the energy constraint. A larger $\mu$ places greater emphasis on code size reduction, whereas a smaller $\mu$ prioritizes execution time reduction.

We later, in Section~\ref{subsec:GNNPP}, connect the optimization objective in Formula~\ref{eq:problem} with the reward function of the RL algorithm.

\section{MileStone Framework}\label{sec:system}

\subsection{Overall Structure}

\begin{figure}[t]
    \centering
    \includegraphics[width=\textwidth]{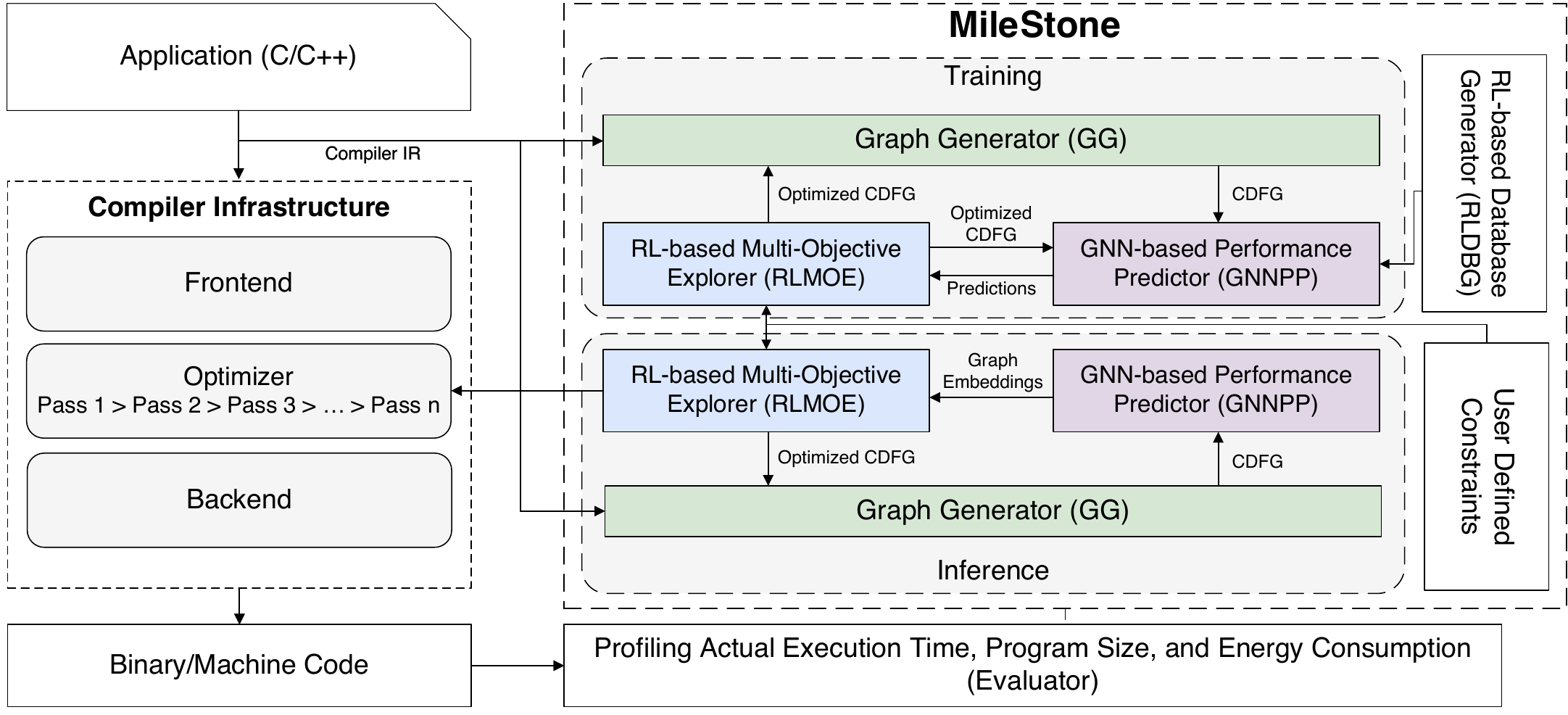}
    \caption{MileStone High-level architectural view.}
    \label{fig:framework}
\end{figure}

Figure~\ref{fig:framework} presents the overall architecture of MileStone, comprising four key modules: (1) Graph Generator (GG), (2) GNN-based Performance Predictor (GNNPP), (3) RL-based Multi-Objective Explorer (RLMOE) and (4) RL-based Database Generator (RLDBG).

\begin{enumerate}
    \item \textit{GG} serves as the interface between GNNPP/RLMOE and compiler, which uncovers hidden optimization opportunities, and enables flexible and fine-grained optimization space exploration under user-defined constraints. It extracts CDFGs after the front-end compilation stage of LLVM to expose additional optimization opportunities.
    \item \textit{GNNPP} is a GNN-based performance predictor that estimates post-compilation metrics, including code size, execution time and energy consumption of CDFGs. It appears in two parts of the MileStone framework. (1) In training phase where it uses RLDBG to generate a dataset for training GNNPP, and receives graphs as CDFG from GG. As an expected output, it produces cumulative reward, and passes it to RLMOE.(2) In inference phase where it receives raw CDFG graphs from GG and it embeds LLVM IR instructions within the graph in the form of graph embedding as the output. 
    \item \textit{RLMOE} is an RL-based optimization space exploration engine. It takes as input the raw CDFG, its graph embedding and user-defined constraints to optimize compiler transformation strategies. RLMOE integrates with GNNPP, where graph embeddings serve as the state representations, capturing the structural and semantic relationships of CDFGs. These embeddings enable RLMOE to generalize effectively across diverse CDFG topologies, while GNNPP accelerates RLMOE’s training process by rapidly providing performance predictions (e.g., code size, execution time and energy consumption) for evaluating intermediate solutions. Consequently, this synergy allows RLMOE to conduct efficient and fine-grained design space exploration with improved generalization and faster convergence.
    
    \item \textit{RLDBG} is a self-evolving knowledge base that aggregates compiler IRs, CDFGs and measured performance metrics to enable supervised training of GNNPP and refinement of RLMOE. It automates large-scale exploration of compiler pass sequences using RLMOE, compiles and profiles each configuration through the Evaluator, and stores the resulting execution time, code size and energy data alongside CDFG metadata. By capturing diverse and Pareto-efficient optimization outcomes, RLDBG eliminates redundant profiling, allowing GNNPP to predict performance metrics efficiently and accelerating the feedback loop within MileStone.
\end{enumerate}

As a case study of MileStone, the specific problem addressed is the phase ordering optimization under strict energy constraints. The objective is to identify an optimization sequence that satisfies the user-specified energy consumption constraint while either (1) minimizing code size within the allowed energy budget or (2) discovering Pareto-optimal trade-offs among energy consumption, code size and execution time, without compromising computational latency.

The work flow of MileStone proceeds through four tightly integrated stages, forming a closed-loop optimization cycle that bridges compiler analysis, learning-based prediction and adaptive exploration. First, the Graph Generator extracts control flow graphs and corresponding metadata from LLVM IR after the front-end compilation stage. Next, the RLDBG module populates a large-scale database by iteratively invoking RLMOE to explore diverse compiler pass sequences. The labeled CDFG–metric pairs serve as training data for GNNPP, which learns to predict post-compilation performance metrics directly from CDFG representations. Once trained, GNNPP provides rapid performance estimation to guide RLMOE, which performs multi-objective optimization under user-defined constraints using a reinforcement learning technique (e.g., PPO or DQN). This integration enables efficient exploration of vast optimization spaces without repeated physical execution. The feedback between RLMOE and GNNPP continuously refines both modules and Ultimately, MileStone identifies Pareto-optimal compiler optimization strategies that balance execution time, code size and energy consumption, achieving high-quality solutions with minimal profiling overhead.

\subsection{Proposed GNNPP and RLMOE}\label{subsec:GNNPP}

\subsubsection{\textbf{GNNPP: GNN-Based Performance Predictor}}

\hfill\\

\textbf{Node Feature Vector.} Each node in the CDFG is represented by a 10-dimensional binary feature vector that encodes both structural and semantic information about the nodes. The first feature distinguishes node types: it is set to 1 for basic block nodes and 0 for instruction nodes. The remaining nine features employ a one-hot encoding scheme to represent the instruction type for instruction nodes, capturing the most common LLVM IR operations: memory allocation (alloca), memory load/store operations (load, store), arithmetic operations (add, sub, mul, div), comparison operations (icmp) and function calls (call). For basic block nodes, all instruction-type features are set to 0, while for instruction nodes, exactly one of the instruction-type features is set to 1 corresponding to the node's opcode, with the remaining features set to 0. This compact representation enables the GNN to distinguish between control flow structures (basic blocks) and computational operations (instructions).

\textbf{Graph Embedding.} We use a graph convolutional network to predict performance metrics of compiled code. The propagation rule of each graph convolutional layer is expressed as:

\begin{equation}
H^{(l+1)} = \sigma \left( \widetilde{D}^{-\tfrac{1}{2}} \times \widetilde{A} \widetilde{D}^{-\tfrac{1}{2}}\times H^{(l)} \times W^{(l)} \right),
\end{equation}

\noindent where $\widetilde{A} = A + I_N$ is the sum of the adjacency matrix $A$ of the CDFG $G$ and the identity matrix $I_N$, ensuring self-loops are included so that each node representation aggregates information from its neighbors and itself. $\widetilde{D}$ is the diagonal degree matrix for normalization, $\widetilde{D}_{ii} = \sum_j \widetilde{A}_{ij}$. $W^{(l)}$ is the layer-specific trainable weight matrix. $\sigma(\cdot)$ is the activation function. $H^{(l)} \in \mathbb{R}^{n \times d}$ represents the matrix of node activations at layer $l$, with $n$ nodes each having feature dimension $d$. $H^{(0)}$ is the initial matrix of input node feature vectors.

\begin{figure}[t]
    \centering
    \includegraphics[width=\textwidth]{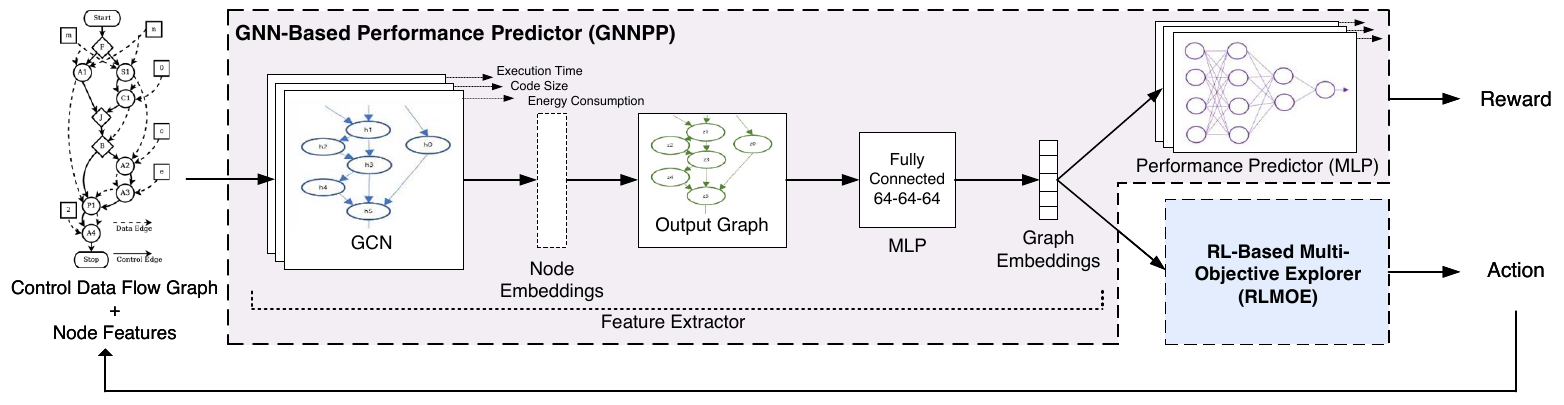}
    \caption{Detailed flow of integrated GNNPP and RLMOE modules.}
    \label{fig:details}
\end{figure}

In our GNNPP design, the inputs to the first GCN layer are the adjacency matrices and node feature matrices of CDFGs. At each layer, embeddings are updated by aggregating information from a node’s neighbors, weighted by $W^{(l)}$. Stacking multiple layers allows capturing multihop dependencies in the CDFG. Figure~\ref{fig:details} illustrates an example GCN with two layers. After several layers, the learned node embeddings are aggregated using mean pooling to produce a graph-level representation.

we define $\alpha = (\mu/q)$ and $\lambda = 1 - \mu$. If additional metrics or alternative forms of multi-objective optimization are needed, the reward function can be extended or redefined accordingly. Furthermore, by adjusting the weights (i.e., hyperparameters) of the selected metrics in Formula~\ref{eq:problem}, our proposed method in Section~\ref{rlmoe} can explore different tradeoffs while meeting user-defined constraints.This representation is then passed to a feedforward network with three fully connected layers and leaky ReLU ($\alpha = 0.1$) activations to generate a graph embedding. The final layer outputs predictions for code size, energy consumption and execution time. To separately predict each of these metrics, we employ three GNN models with the same architecture, as shown in Figure~\ref{fig:details}.

\textbf{Integration with RLMOE.} To integrate GNNPP with RLMOE, the three embedding vectors (corresponding to code size, energy consumption and execution time) are concatenated into a single representation vector of size $192 \times 1$, as illustrated by Figure~\ref{fig:details}. This vector is then concatenated with metadata of the input CDFG, such as the number of input/intermediate/output nodes, number of edges, and number of multiplications. The combined representation is fed to RLMOE as input.  

Given GNNPP’s predictions of code size, energy consumption and execution time, RLMOE can generate optimized solutions, which are quickly evaluated and used as feedback to further refine the policy of RLMOE.

\subsubsection{\textbf{RLMOE: RL-Based Multi-Objective Explorer}}
\label{rlmoe}

\hfill\\

The phase order space exploration in compiler optimization can be modeled as an RL task, expressed as an MDP with four primary elements:

\begin{enumerate}
    \item \textbf{States:} The set of all possible states. In this problem, a state corresponds to a partially assigned CDFG.
    \item \textbf{Actions:} The set of valid actions under a given state. At each step, given the current state and the currently considered node of the CDFG, the action determines whether a directive should be applied to that node.
    \item \textbf{State Transition:} Defines the probability distribution of next states given the current state and chosen action.
    \item \textbf{Reward:} It shows the immediate reward for taking an action. In this case, all intermediate rewards are zero, except for the final step, where the reward evaluates the fully assigned CDFG under user-specified constraints.
\end{enumerate}

At time step $t$, the state $s_t$ is defined as a feature vector including a $192 \times 1$ graph embedding that describes the current CDFG status, the identifier of the current node, metadata of the CDFG, and the energy consumption constraint (either user-defined or automatically generated for Pareto exploration). The action $a_t$ represents the assignment of a directive to the $t$-th node, e.g., whether to optimize for code size or execution time at that point.  

The formulas and notations we use in this section are inspired by the main references of DQN and PPO techniques~\cite{DQN1, ppo1}. The reward $r_t$ is expressed as a negatively weighted sum of predicted code size, execution time and the difference between predicted and target energy consumption values, which naturally supports multi-objective optimization:

\begin{equation}\label{eq:rewards}
r_t = 
\begin{cases} 
-\alpha \text{CodeSize}_p - \beta \lvert \text{Energy}_t - \text{Energy}_p \rvert - \lambda \text{ExecTime}_p, & t = T \\
0, & 0 < t < T
\end{cases}
\end{equation}

\noindent where $\alpha$, $\beta$ and $\lambda$ are hyperparameters; $\text{Energy}_t$ is the target energy consumption; and $\text{Energy}_p$, $\text{CodeSize}_p$ and $\text{ExecTime}_p$ are predicted values from the performance model. $T$ denotes the total number of steps.

At the initial state $s_0$, all CDFG nodes are unassigned. At each step, the RL agent observes the state $s_t$, selects an action $a_t$, receives a reward $r_{t+1}$, and transitions to $s_{t+1}$. Nodes are sequentially assigned directives based on their IDs until the final state $s_T$, where the CDFG is fully annotated. The goal is to maximize expected cumulative rewards.

\textbf{Training.} RLMOE supports two RL methods, providing flexibility in handling different optimization scenarios.

\noindent \textbf{(1) DQN Method:} We adopt the Deep Q-Network approach, where a Q-function $Q(s,a;\theta)$ is approximated by a neural network with parameters $\theta$, serving as an estimator of the optimal action-value function:
    \begin{equation}
    Q(s,a;\theta) \approx Q^*(s,a) = \max_\pi \mathbb{E}_\pi \left[ \sum_{t=0}^{\infty} \gamma^t r_t \; \Big| \; s_0=s, a_0=a \right],
    \end{equation}
    where $\gamma \in [0,1]$ is the discount factor and $\pi$ denotes a policy.  
    
    The network parameters are optimized by minimizing the temporal-difference (TD) loss:
    \begin{equation}
    L(\theta) = \mathbb{E}_{(s,a,r,s') \sim \mathcal{D}} \Big[ \big( y^{\text{DQN}} - Q(s,a;\theta) \big)^2 \Big],
    \end{equation}
    where $\mathcal{D}$ is the replay buffer and the target $y^{\text{DQN}}$ is computed using a target network $Q(s,a;\theta^-)$ with periodically updated parameters $\theta^-$ as:
    \begin{equation}
    y^{\text{DQN}} = r + \gamma \max_{a'} Q(s',a';\theta^-).
    \end{equation}
    
    The gradient update for $\theta$ is performed via stochastic gradient descent:
    \begin{equation}\label{eq:rlmoe-dqn}
    \theta \leftarrow \theta - \alpha \nabla_\theta L(\theta),
    \end{equation}
    where $\alpha$ is the learning rate. To stabilize learning, the target network parameters are updated every $C$ iterations as $\theta^- \leftarrow \theta$.
    
\noindent \textbf{(2) PPO Method:} Alternatively, we employ Proximal Policy Optimization (PPO), a policy-gradient method designed to enhance training stability through clipped updates. Given the MDP formulation and reward $r_t$ defined in Formula~\ref{eq:reward}, the policy $\pi_\theta(a|s)$ is optimized by maximizing the clipped surrogate objective:
    \begin{equation}\label{eq:reward}
    L^{\text{PPO}}(\theta) = \mathbb{E}_t \Big[ \min \big( \rho_t(\theta) \hat{A}_t, \; \text{clip}(\rho_t(\theta), 1-\epsilon, 1+\epsilon) \hat{A}_t \big) \Big],
    \end{equation}
    where $\rho_t(\theta) = \frac{\pi_\theta(a_t|s_t)}{\pi_{\theta_\text{old}}(a_t|s_t)}$ is the probability ratio between new and old policies, 
    $\hat{A}_t$ is the advantage estimate, and $\epsilon$ is a clipping threshold that constrains the policy update.
    
    The advantage $\hat{A}_t$ is computed using the Generalized Advantage Estimation (GAE)~\cite{GAE1}:
    \begin{equation}
    \label{eq:eq10}
    \hat{A}_t = \sum_{l=0}^{\infty} (\gamma \lambda)^l \, \delta_{t+l},
    \end{equation}
    where the temporal-difference residual is defined as
    \begin{equation}
    \delta_t = r_t + \gamma V(s_{t+1};\phi) - V(s_t;\phi),
    \end{equation}
    and $V(s_t;\phi)$ denotes the value function with parameters $\phi$.
    The value network is trained to minimize:
    \begin{equation}
    L^{\text{VF}}(\phi) = \mathbb{E}_t \Big[ \big( V(s_t;\phi) - V_t^{\text{target}} \big)^2 \Big],
    \end{equation}
    with the target value $V_t^{\text{target}} = r_t + \gamma V(s_{t+1};\phi)$.
    
    The final PPO objective combines the policy loss, value loss, and entropy regularization term to encourage exploration:
    \begin{equation}
    L^{\text{total}}(\theta, \phi) = \mathbb{E}_t \Big[ L^{\text{PPO}}(\theta) - c_1 L^{\text{VF}}(\phi) + c_2 H(\pi_\theta(s_t)) \Big],
    \end{equation}
    where $H(\pi_\theta(s_t)) = - \sum_a \pi_\theta(a|s_t) \log \pi_\theta(a|s_t)$ is the policy entropy, and $c_1$, $c_2$ are weighting coefficients.
    
    The gradient update proceeds via gradient ascent:
    \begin{equation}
    \label{eq:eq14}
    \theta \leftarrow \theta + \alpha_\pi \nabla_\theta L^{\text{PPO}}(\theta)
    \end{equation}
    \begin{equation}
    \label{eq:eq15}
    \phi \leftarrow \phi - \alpha_V \nabla_\phi L^{\text{VF}}(\phi).
    \end{equation}
    where $\alpha_\pi$ and $\alpha_V$ are the learning rates for the policy and value networks, respectively.

\begin{algorithm}[t]
\caption{DQN/PPO for Phase Ordering Optimization}
\label{alg:rlmoe-dqn-ppo}
\begin{algorithmic}[1]
\State Create tuples $\big[\textit{CDFG}_{\text{index}}, \textit{Energy}_{\text{target}}\big]$, duplicate each $m$ times, then shuffle;
\State \textbf{Initialize:}
\State \quad \textit{DQN:} online Q-network $Q(s,a;\theta)$, target network $Q(s',a';\theta^-)$;
\State \quad \textit{PPO:} policy network $\pi_\theta(a|s)$ and value $V_\phi(s)$;
\State \quad Episode counter $i \gets 0$;
\While{$i < \textit{episode}_{\max}$}
    \State Get the $\textit{CDFG}_{\text{index}}$ and $\textit{Energy}_{\text{target}}$;
    \State $t \gets 0$;
    \State Build initial state $s_0$ with $\textit{CDFG}_{\text{index}}$ embedding and metadata;
    \While{$t < T$}
        \State \textit{DQN:} $a_t \gets \varepsilon$-greedy over $Q(s,\cdot;\theta)$;
        \State \textit{PPO:} sample $a_t \sim \pi_\theta(\cdot|s_t)$ and record $\log \pi_\theta(a_t|s_t)$ and $V_\phi(s_t)$;
        \State Compute reward $r_{t+1}$ as defined in Formula~(\ref{eq:rewards});
        \State Form next state $s_{t+1}$ as updated CDFG embedding and metadata;
        \State $t \gets t+1$;
    \EndWhile
    \State \textit{DQN:} update $\theta$ and $\theta^-$ according to Formula~(\ref{eq:rlmoe-dqn});
    \State \textit{PPO:} compute advantage and update $\theta$ and $\phi$ according to Formulas~(\ref{eq:eq10}),~(\ref{eq:eq14}) and~(\ref{eq:eq15});
    \State $i \gets i+1$;
\EndWhile
\end{algorithmic}
\end{algorithm}

Algorithm~\ref{alg:rlmoe-dqn-ppo} describes how the RLMOE module trains either a DQN or a PPO agent to assign optimization directives to the nodes of a CDFG under a given energy constraint. It begins in line~1 by generating tuples of CDFG indices and energy targets, and it increases the number of samples by duplicating each tuple. Line~2 shuffles these tuples to avoid training bias. Lines~3--4 initialize the required learning models: the online and target Q-networks 
for DQN, or the policy and value networks for PPO. The episode counter is set in line~5. Lines~6--7 start the main training loop and retrieve the CDFG and its target energy value for the current episode. Line~9 constructs the initial state, which includes the graph embedding, CDFG metadata and the energy constraint.

Lines~10--16 perform the step-by-step interaction between the agent and the CDFG. At each step, line~11 selects an action using an $\epsilon$-greedy rule when DQN is used, while line~12 samples 
an action from the PPO policy and stores its log-probability and value estimate. Line~13 computes the reward, which is zero for all intermediate steps and non-zero only at the final step, based on 
predicted performance metrics. Line~14 builds the next state by updating the embedding and metadata. After all nodes are processed, line~17 updates the DQN parameters using TD learning, and line~18 updates the PPO policy and value networks using the clipped surrogate objective. Line~19 increases the episode counter, and the loop continues until the maximum number of episodes is reached.

\textbf{Generalization Across CDFGs.} The objective of RLMOE is to produce high-quality solutions and transfer knowledge across CDFGs as it gains experience. The overall optimization objective is defined as:

\begin{equation}
\mathcal{J}(\theta, w, G) = \frac{1}{K} \sum_{g \in G} \mathbb{E}_{p \sim \pi_\theta}[R_{g,p}],
\end{equation}

where $\mathcal{J}(\theta, w, G)$ measures the expected cumulative rewards over all training CDFGs. The dataset $G$ contains $K$ CDFGs, each denoted as $g$. $R_{g,p}$ is the episodic reward (i.e., $r_T$) under optimization strategy $p$ for CDFG $g$. To encourage exploration during training, an $\epsilon$-greedy strategy is applied.  

\textbf{Fine-Tuning.} When presented with a new CDFG, the pretrained RLMOE can be used directly for inference, generating solutions within seconds. For higher-quality solutions, the pretrained model can be fine-tuned on the specific CDFG. This fine-tuning balances the trade-off between fast inference (leveraging prior knowledge from other CDFGs) and extended optimization for improved results.

\subsection{Proposed the RLDBG Module}

\begin{figure}[t]
    \centering
    \includegraphics[width=\textwidth]{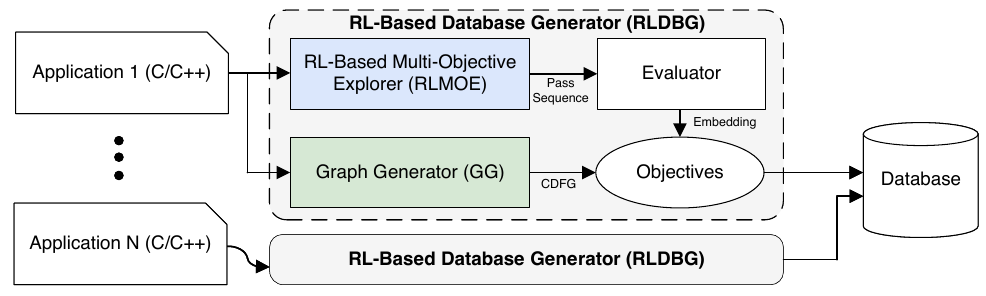}
    \caption{Overview of proposed RLDBG.}
    \label{fig:rldbg}
\end{figure}

As shown by Figure~\ref{fig:rldbg}, the \textit{RLDBG} module is designed to populate a comprehensive database of compiler intermediate representations and their corresponding optimization outcomes, enabling supervised training of GNNPP. It accepts a collection of C/C++ applications as input and leverages RLMOE to iteratively explore various compiler pass sequences and transformation strategies. Each candidate configuration is compiled, profiled, and analyzed by the Evaluator, which measures key performance indicators such as execution time, code size and energy consumption. These results are stored in the database alongside the corresponding CDFGs generated by GG.

For each input application, RLDBG operates as follows:
\begin{enumerate}
    \item The source code is compiled into LLVM IR, and the CDFG is extracted by the Graph Generator.   
    \item The RLMOE is employed to navigate the vast optimization space of compiler passes for each CDFG. It iteratively applies sequences of LLVM passes guided by multi-objective reward functions considering execution time, energy consumption, and code size. By learning from previous exploration episodes, RLMOE efficiently discovers diverse and near-optimal optimization sequences, ensuring the resulting database captures a broad range of trade-offs and Pareto-efficient configurations.
    \item For every optimization sequence discovered by RLMOE, the corresponding compiled binary is executed by the Evaluator, to profile and obtain ground-truth metrics (e.g., execution time, program size, and energy usage), which are essential for supervised training of GNNPP.
    \item The measured performance metrics, combined with graph representations and metadata generated by GG, are stored as labeled entries in the database.  
\end{enumerate}

Finally, RLDBG aggregates all collected CDFGs, optimization sequences, and evaluated objectives into a comprehensive Database, which forms the foundation for training GNNPP to predict compiler optimization outcomes without repeated physical execution, dramatically reducing profiling costs and enabling faster inference during the exploration stage of MileStone.

\section{Experiment}\label{sec:exp}

\subsection{Experimental Setup}

To evaluate the generalization capability of MileStone across diverse control flow graphs and applications, both GNNPP and RLMOE are trained on a subset of CDFGs and tested on the remaining ones. The training dataset comprises CDFGs extracted from PolyBench benchmarks~\cite{polybench}, with three benchmarks (\textit{gemm}, \textit{3mm}, and \textit{floyd-warshall}) reserved for inference to assess generalization performance.

GNNPP is trained as a regression model to minimize the mean-squared logarithmic error for energy consumption and code size, and the mean absolute error for execution time. It is optimized using Adam~\cite{adam} with an exponentially decaying learning rate initialized at 0.01, trained over 200 epochs with a batch size of 32. After training phase, GNNPP assists RLMOE by providing accurate performance predictions during the reinforcement learning process. RLMOE is trained using tuples of the form $(\text{CDFG}_{\text{index}}, \text{Energy}_{\text{target}})$, where the goal is to maximize the cumulative reward across all tuples to learn effective performance strategies under various energy constraints. The training set includes 1050 unique tuples, each revisited eight times, resulting in 3000 episodes in total. After training, RLMOE can rapidly infer optimized allocation strategies for unseen CDFGs.

RLMOE parameters are optimized via Adam with a learning rate of 0.008. The discount factor and exploration rate are set to $\gamma = 0.95$ and $\epsilon = 0.08$, respectively, with exponential decay. In the reward formulation, $\beta = 5$ and $q = 500$, with three trade-off scenarios between execution time and code size defined by $\mu \in \{0.1, 0.5, 0.9\}$.

The MileStone experiments are evaluated using \textsc{StellarGraph} for GNNPP and TensorFlow~2 for RLMOE. All experiments are conducted on a Linux workstation equipped with an Nvidia RTX~3090TI GPU, an Intel Core i7-1165G7 CPU and 24GB RAM.

\subsection{Evaluation}

\subsubsection{Baselines}

To evaluate the effectiveness of the proposed MileStone framework, we compare the performance of its components against a diverse set of baselines that include traditional compiler optimization levels, heuristic search algorithms, standalone RL approaches and related learning-based frameworks for compiler phase ordering.

\textbf{LLVM Built-in Optimization Levels.} We first consider LLVM’s canonical optimization levels -O1, -O2, and -O3, which serve as standard baselines in compiler optimization research.
These predefined pass sequences represent heuristics tuned by compiler engineers to balance compilation time and performance.
While -O1 applies conservative transformations for faster compilation, -O2 introduces a broader set of optimizations for balanced performance, and -O3 enables the most aggressive suite of optimizations, such as loop unrolling, inlining, and auto-vectorization, aimed at maximizing execution speed.

\textbf{Heuristic Search Methods.} Two population-based metaheuristic algorithms are used as learning-free search baselines:

\begin{itemize}
    \item Genetic Algorithm (GA)~\cite{phase8}: evolves candidate phase-ordering sequences through crossover and mutation. GA assumes that offspring derived from strong individuals inherit superior traits; however, this assumption often fails in compiler optimization spaces with weak structural inheritance between passes.
    \item Particle Swarm Optimization (PSO)~\cite{phase2}: explores the search space by iteratively adjusting candidate solutions based on collective experience. Although computationally efficient, PSO lacks evolutionary operators and tends to converge slowly or become trapped in local optima in high-dimensional, irregular landscapes.
\end{itemize}

\textbf{Reinforcement Learning Baselines.} We further include Deep Q-Network (DQN) and Proximal Policy Optimization (PPO) as standalone RL agents that learn compiler pass sequences without the full multi-objective modeling of RLMOE.
These methods are trained directly on the performance metrics (execution time, code size, and energy consumption) extracted via RLDBG.

\textbf{Related Learning-Based Frameworks.} To provide context among learning-based compiler optimization systems, we further compare with POSET-RL~\cite{phase5}, a reinforcement-learning-driven phase-ordering framework that models the compiler optimization process as a partially ordered set (POSET) of transformations from LLVM -Oz optimizion level.
POSET-RL uses Double DQN (DDQN) as RL agent, and IR2Vec encoding for representing programs. However, unlike MileStone, it does not incorporate multi-objective optimization or graph-based program representations, limiting its ability to balance competing metrics such as code size and energy consumption. As a result, we have used a same reward function with POSET-RL and also adopt POST-RL to use -O3 optimization passes to conduct fair comparison.

\subsubsection{\textbf{RLDBG Evaluation}}

\begin{figure}[t]
    \centering
    \includegraphics[width=\textwidth]{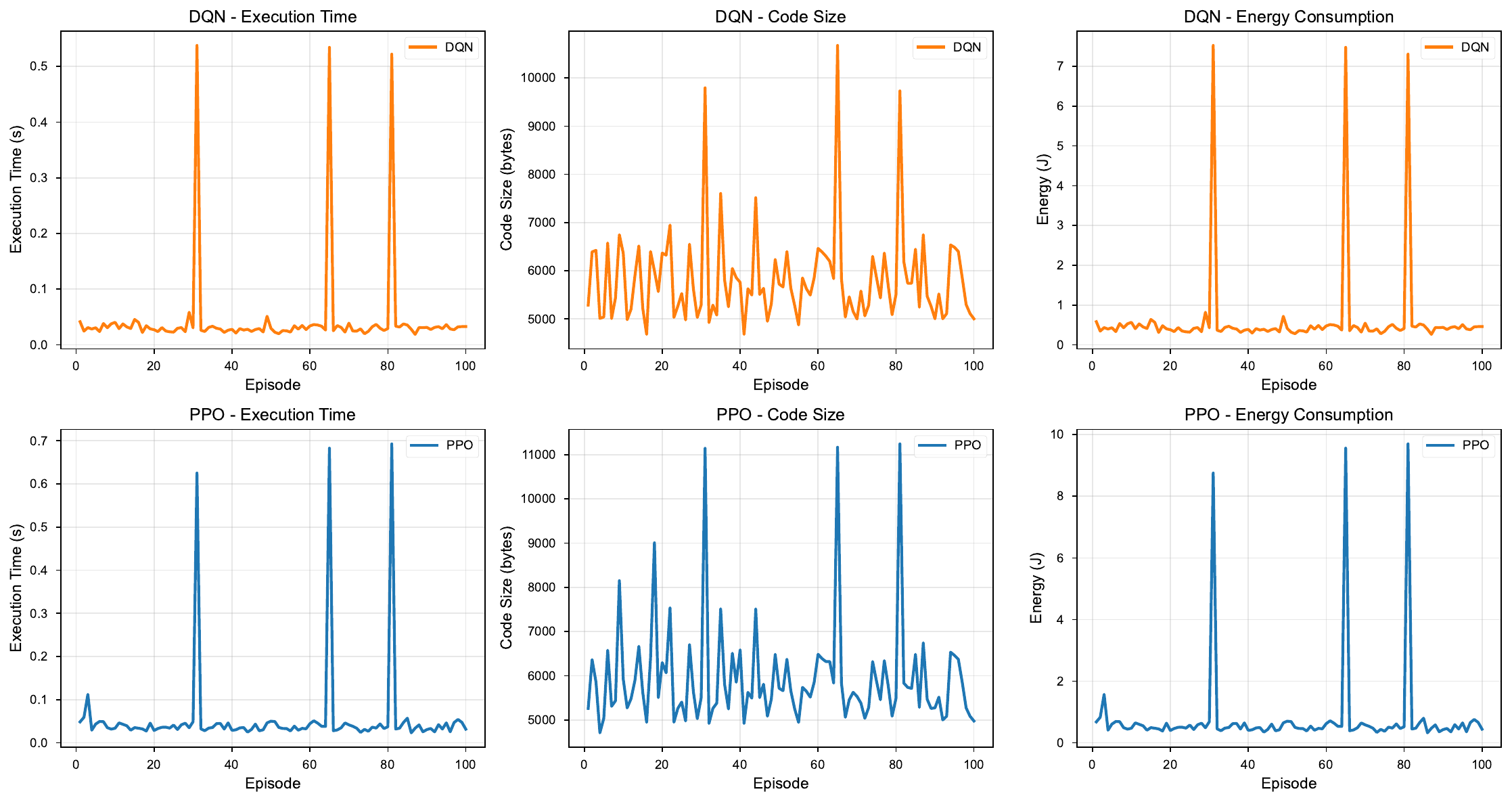}
    \caption{Profiling Execution Time, Code Size, and Energy Consumption for training RLMOE agents (DQN and PPO) equipped with RLDBG over 100 episodes.}
    \label{fig:agent}
\end{figure}

To evaluate the proposed RL-Based Database Generator, we conducted experiments using RLMOE integrated within RLDBG, as illustrated in Figure~\ref{fig:rldbg}. Profiling results for two reinforcement learning agents (DQN and PPO) are summarized in Figure~\ref{fig:agent}. Each agent is trained for 100 episodes under identical RLDBG configurations to assess execution time, code size and energy consumption trends. Both agents exhibit periodic spikes in execution time and energy usage, typically corresponding to exploration-intensive episodes or high-complexity code transformations. As denoted by Figure~\ref{fig:agent}, DQN demonstrated slightly higher variance across all three metrics, while PPO maintained more stable execution and energy efficiency. These observations highlight that RLDBG effectively captures the dynamic trade-offs between performance and resource consumption across different RL-based optimization strategies, validating its suitability for constructing rich, multi-objective optimization databases.

\subsubsection{GNNPP Evaluation}

\begin{figure}[t]
    \centering
    \includegraphics[width=\textwidth]{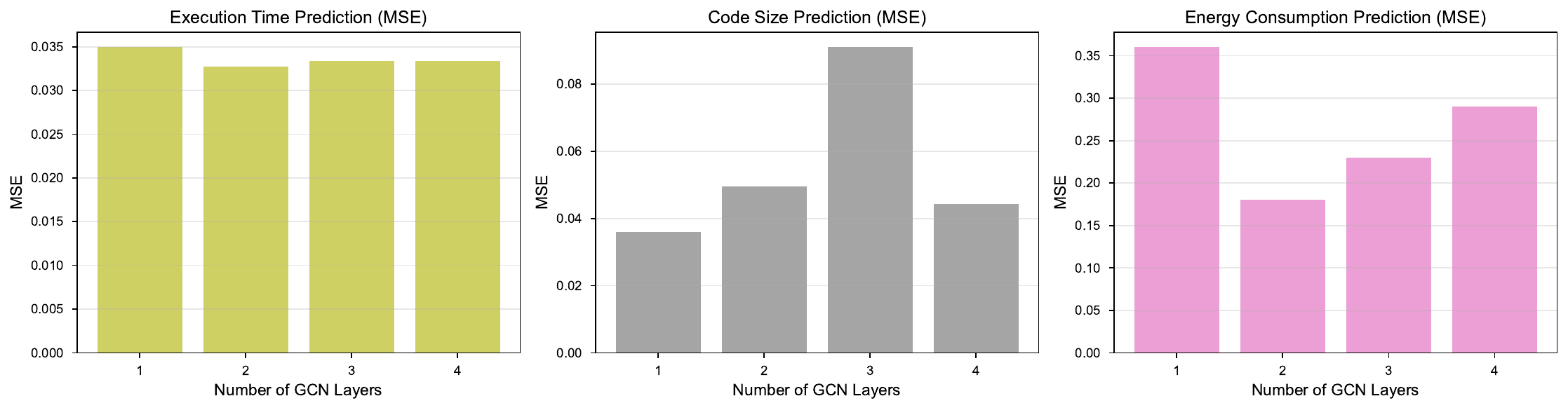}
    \caption{Comparison the prediction accuracy of GNNPP in MSE, where applying different numbers of GCN layers.}
    \label{fig:GNNPP-eval}
\end{figure}

Excessive stacking of graph convolutional layers often leads to the oversmoothing problem~\cite{oversmooth-graph}, where node features become indistinguishable across the graph. To determine an effective architecture for GNNPP, we systematically vary the number of GCN layers used for feature extraction and compare their predictive performance. As denoted by Figure~\ref{fig:GNNPP-eval}, the configuration employing two GCN layers achieves consistently promising accuracy on predicting execution time, code size and energy consumption (with the actual performance of program after optimization as a grand truth) compared to deeper variants. Hence, this two-layer design is adopted as the final architecture of GNNPP. This analysis reveals that increasing model depth does not always improve performance, and careful architecture selection is crucial for achieving optimal prediction accuracy across different program characteristics.

\subsubsection{RLMOE and MileStone Evaluation}

We evaluate the effectiveness of MileStone from two complementary perspectives: (1) Pareto-optimal solutions, and (2) user-specified constraints.

\begin{figure}[t]
    \centering
    \includegraphics[width=\textwidth]{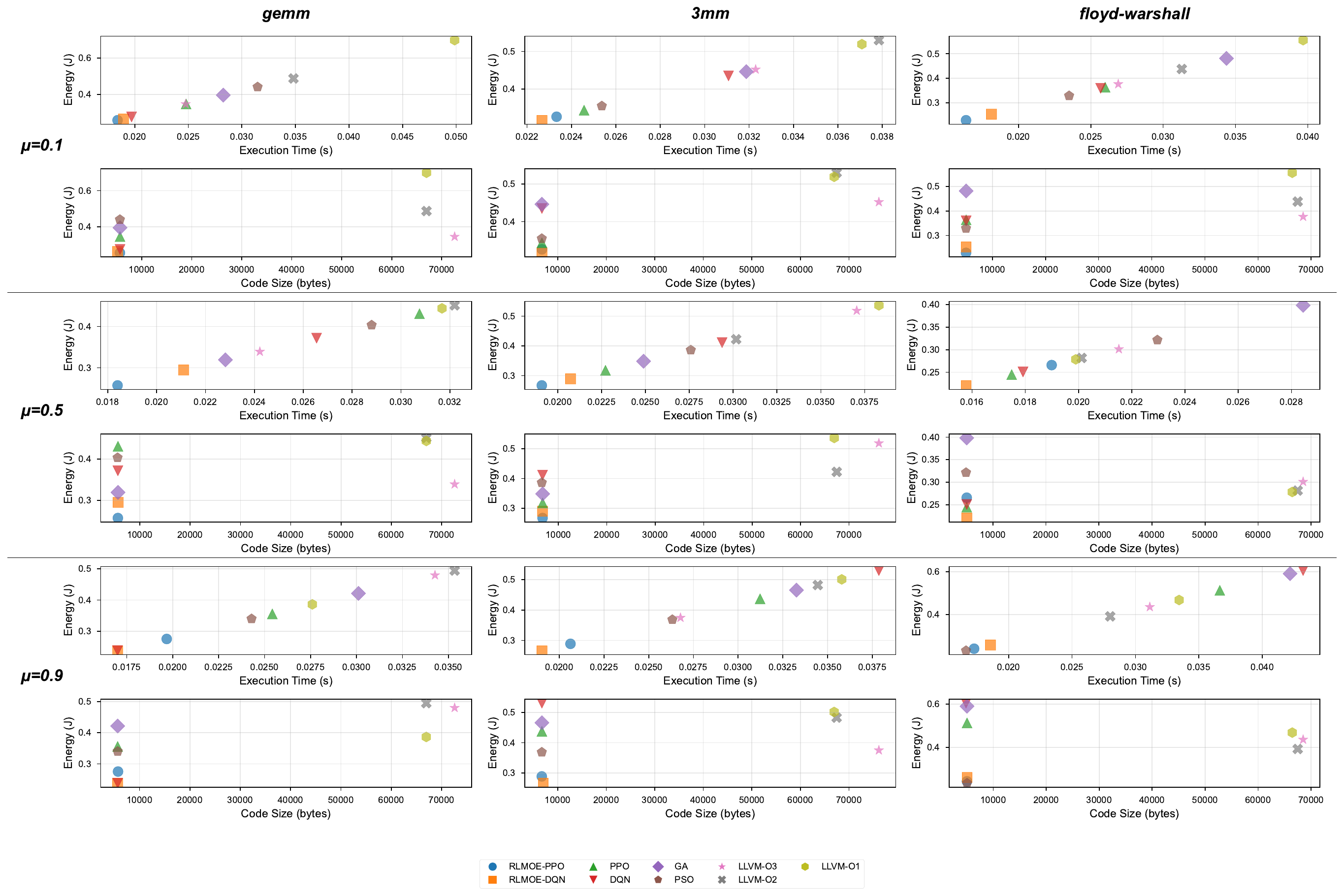}
    \caption{The Actual inference performace gained by RLMOE (either with DQN or PPO as agent), PPO, DQN, GA, PSO, -O1, -O2, and -O3 optimizations on pass sequences for three real case benchmarks, gemm, kernel\_3mm, and Floyd-Warshall targeting minimum energy consumption. Different $\mu$ values indicate different importance of execution time and code size objectives in optimization.}
    \label{fig:2d}
\end{figure}

\textbf{Pareto Solutions.}
Regarding the Pareto solutions between execution time or code size versus energy consumption, Figure~\ref{fig:2d} compares RLMOE with DQN, PPO, GA, PSO and O1/O2/O3 across three benchmarks, respectively. Clearly, RLMOE, whether using DQN or PPO methods, outperforms other methods by a notable margin. In terms of multi-objective optimization, given energy consumption constraints, the solutions with $\mu = 0.9$ often achieve lower execution time but higher (worse) code size compared to those with $\mu = 0.1$. This demonstrates that RLMOE can effectively balance execution time and code size when varying importance is assigned to different objectives. Heuristic-based methods cannot explicitly model such trade-offs.

\begin{figure}[t]
    \centering
    \includegraphics[width=\textwidth]{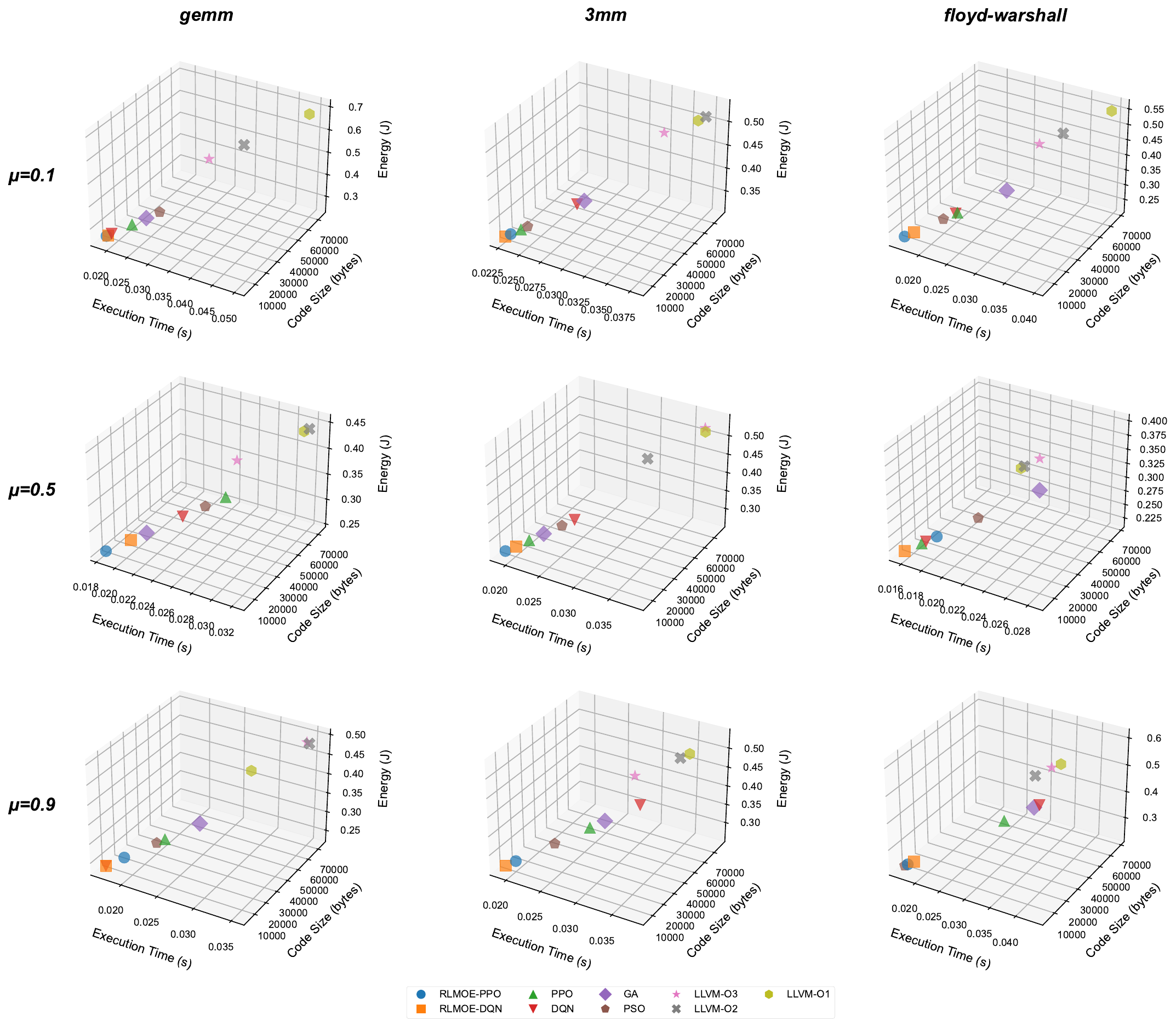}
    \caption{Pareto solutions between three main objectives found by RLMOE, PPO, DQN, GA, PSO, -O1, -O2, and -O3, on three benchmarks, gemm, kernel\_3mm, and Floyd-Warshall.}
    \label{fig:3d}
\end{figure}

Figure~\ref{fig:3d} extends the analysis of Figure~\ref{fig:2d} by presenting a three-dimensional visualization of the Pareto-optimal solutions, simultaneously displaying the trade-offs between execution time, code size, and energy consumption. This three-dimensional representation provides a comprehensive view of the optimization space, allowing for better understanding of how different methods navigate the complex multi-objective landscape. The 3D scatter plots reveal that RLMOE methods consistently occupy regions of the solution space characterized by lower energy consumption, smaller code sizes, and competitive execution times. This forms distinct clusters that demonstrate their superior optimization capabilities. The visualization across different $\mu$ values (0.1, 0.5, 0.9) shows how the weighting parameter influences the distribution of solutions in the three-dimensional space, with lower $\mu$ values pushing solutions toward lower energy consumption regions, while higher $\mu$ values allow for more balanced distributions. The 3D perspective highlights the advantage of RLMOE approaches in finding solutions that simultaneously optimize all three objectives, rather than trading off between pairs of metrics, denoting the effectiveness of the multi-objective reinforcement learning framework in discovering superior optimization strategies.

Comparing DQN and PPO, PPO generally provides better outcomes, especially for large-scale programs and complex applications. This is because as the program size increases, the critic network in DQN struggles to precisely estimate state-value functions, hindering policy convergence. Notably, since PPO-generated solutions do not always dominate those from DQN, both are integrated into the MileStone framework to make RLMOE a more capable and adaptive optimization engine.

The promising outcomes denote the strong potential of applying RL for phase ordering space exploration in compiler optimization. Through iterative interactions with GNNPP and user-defined constraints, RLMOE learns which compiler directive should be applied to which node and progressively develops robust allocation strategies through balanced exploration and exploitation. In contrast, GA assumes that offspring from strong individuals are likely superior, an assumption that fails in compiler optimization contexts, reducing its effectiveness. Similarly, PSO lacks evolutionary operators (e.g., crossover, mutation) and tends to converge slowly or get trapped in local optima, especially in high-dimensional spaces. -O3 relies heavily on random exploration, disregarding prior experience, which limits its reliability and consistency.

\begin{figure}[t]
    \centering
    \includegraphics[width=\textwidth]{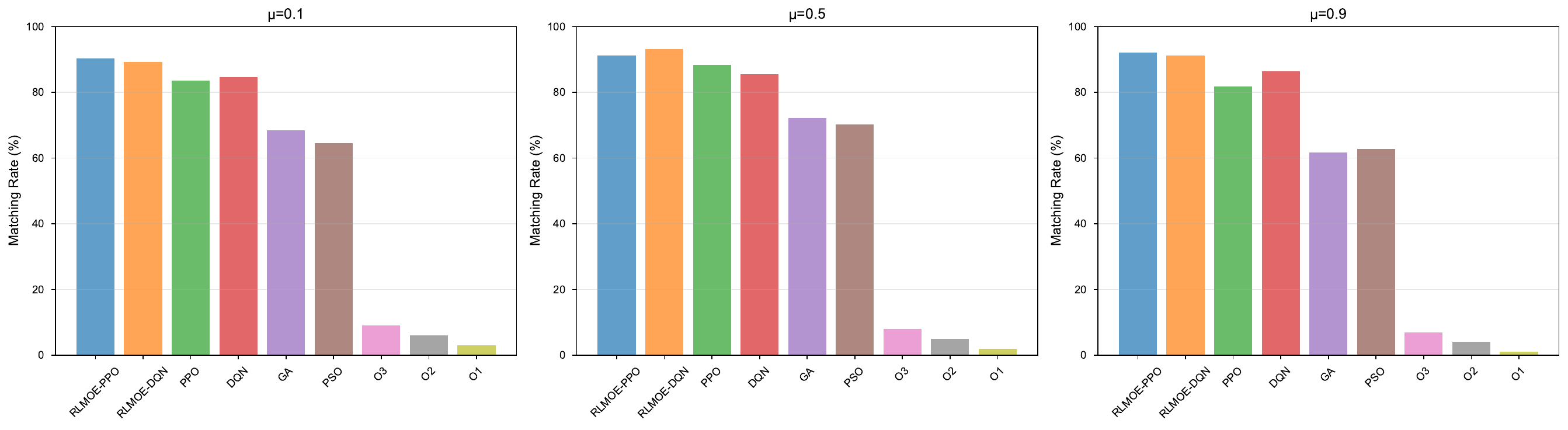}
    \caption{The matching rate of discrete energy consumption constraints is evaluated across GA, PSO, -O1, -O2, -O3, DQN, PPO, and RLMOE (which utilizes either DQN or PPO). Three different settings of $\mu$ are examined, each involving 112 discrete energy consumption constraints applied to three benchmarks: gemm, kernel\_3mm, and floyd-warshall. The reported average matching rate for each method is computed as the arithmetic mean over all 336 constraints.}
    \label{fig:matching}
\end{figure}

\textbf{Solutions Under User-Specifsied Constraints.} To further demonstrate that MileStone effectively satisfies user-specified energy constraints without sacrificing performance, we define discrete energy limits for evaluation. Considering three $\mu$ values and four real-world benchmarks per configuration, yielding 336 total energy constraints, Figure~\ref{fig:matching} depicts RLMOE methods outperform all baseline approaches in consistently meeting energy targets. At $\mu=0.1$, RLMOE-PPO achieves a matching rate of 90.2\%, followed by RLMOE-DQN at 89.3\%, while standalone PPO and DQN achieve 83.6\% and 84.5\%, respectively. Traditional optimization methods (GA, PSO) show lower matching rates of 68.4\% and 64.6\%, while LLVM optimization levels (O3, O2, O1) achieve only 9\%, 6\%, and 3\%, respectively. It indicates their inability to effectively target specific energy constraints. At $\mu=0.5$, RLMOE-PPO reaches 91.2\% matching rate, with RLMOE-DQN achieving 93.1\%, which denotes the effectiveness of the multi-objective framework at intermediate energy-weighting values. At $\mu=0.9$, RLMOE-PPO maintains the highest matching rate at 92.1\%, while RLMOE-DQN achieves 91.2\%. The results consistently show that RLMOE methods achieve matching rates 5--10 percentage points higher than standalone RL approaches and 20--30 percentage points higher than traditional optimization methods. It validates the advantage of integrating graph neural network predictions with RL for energy-constrained optimization scenarios.

\begin{figure}[t]
    \centering
    \includegraphics[width=\textwidth]{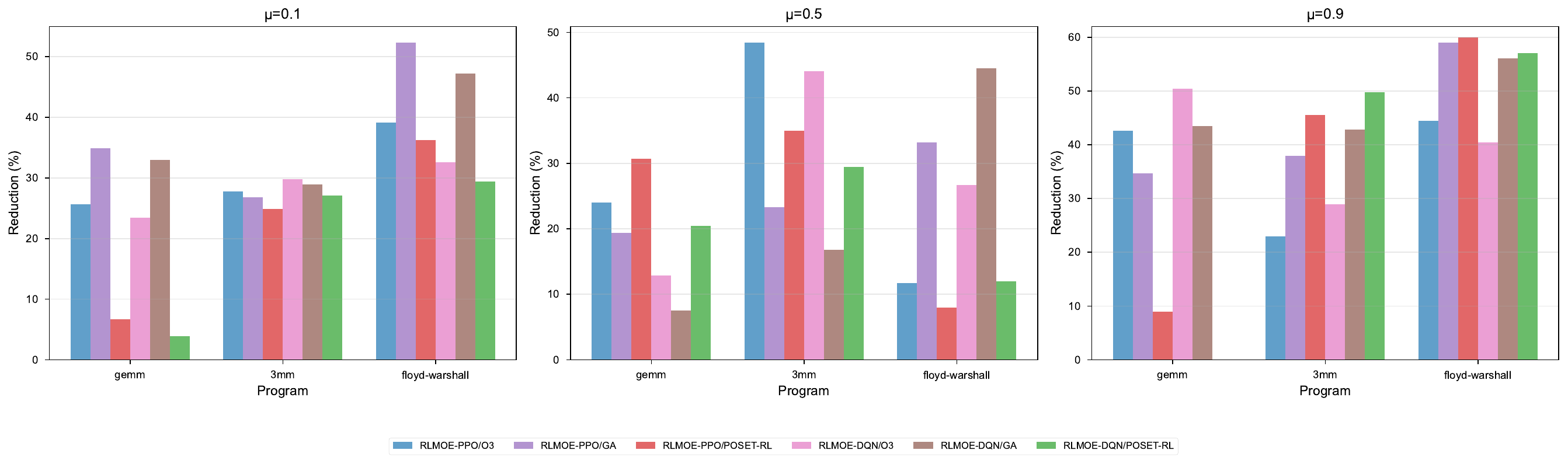}
    \caption{Analysis of reduction in Exection Time given the same Energy Consumption, comparing RLMOE (which applies DQN or PPO) with -O3, GA, and DQN on three real-case benchmarks: gemm, kernel\_3mm, and Floyd-Warshall.}
    \label{fig:reduction}
\end{figure}

Figure~\ref{fig:reduction} shows the statistical results of execution time reduction achieved by MileStone compared to baseline optimization approaches, under identical energy budgets. The results reflect that RLMOE methods consistently achieve execution time reductions across all tested configurations. Specifically, RLMOE-PPO shows reductions ranging from approximately 15-45\% compared to LLVM-O3, with strong performance at $\mu=0.5$ where it achieves reductions of 20-35\% across different benchmarks. When compared to POSET-RL, RLMOE-PPO achieves a 9\% reduction for gemm at $\mu=0.9$ and an 8\% reduction for floyd-warshall at $\mu=0.5$, which illustrate consistent improvements over the baseline reinforcement learning approach. RLMOE-DQN also shows competitive performance, with reductions typically in the range of 10-30\% compared to traditional optimization methods. The figure reveals that the effectiveness of RLMOE methods varies with the $\mu$ parameter, with optimal performance often observed at intermediate $\mu$ values (0.5), where the multi-objective optimization effectively balances energy constraints with execution time minimization. These results validate the superiority of the RLMOE framework, which leverages graph neural network predictions to guide reinforcement learning, over both traditional compiler optimizations and standalone reinforcement learning approaches.

\section{Related Work and Comparison}\label{sec:rel}
\subsection{Related Work}
Modern compilers have modular structure and are easy to modifications for possible optimization via learning-based approaches that enhances the compiler capabilities to automate the optimization process. Researchers introduce a deep learning cost model capable of handling programs with dynamically varying sizes by designing a novel tree-structured program representation and a recursive neural architecture~\cite{comp1}. The code2vec framework proposes a neural architecture that represents code snippets through aggregated abstract-syntax-tree paths, which enables effective learning of semantic code embeddings~\cite{comp2}. It achieves strong predictive performance in method-name prediction and delivers general-purpose applicability for downstream program understanding tasks. COBAYN develops a Bayesian-network–based autotuning framework that models statistical relations between application features and compiler optimizations to guide efficient design-space exploration~\cite{comp3}. Another related work formulates the static phase-ordering problem as a reinforcement learning task, training an agent to select optimization passes using only static IR features~\cite{comp4}. The learned agent outperforms LLVM’s O3 sequence on several benchmarks. ML2Tuner also introduces a multi-level machine-learning autotuner that combines validity prediction and backend-aware performance modeling to avoid invalid configurations and accelerate tuning~\cite{comp5}. BaCO presents a Bayesian-optimization autotuning framework supporting heterogeneous parameter types, explicit and hidden constraints, and highly irregular compiler search spaces~\cite{comp6}. It achieves expert-level performance across CPU, GPU and FPGA compilers with significantly fewer evaluations than prior autotuners.

In terms of graph-based optimization techniques, researchers propose the heterogeneous program graph, which augments abstract syntax trees with explicit node and edge types to overcome the ambiguity inherent in homogeneous graph representations~\cite{graph1}. The approach achieves more precise semantic modeling and surpasses state-of-the-art baselines on method-name prediction and code-classification tasks. Another research constructs rich program graphs that integrate syntactic relations with semantic information such as data flow and type hierarchies, which enable Gated Graph Neural Networks to reason over long-range dependencies in code~\cite{graph2}. A related work introduces a general representation based on paths extracted from abstract syntax trees, allowing learning algorithms to leverage structural relationships among program elements without manual feature engineering~\cite{graph3}. The path-based representation generalizes across languages and tasks. The IR2Vec project generates hierarchical vector embeddings of programs by combining learned LLVM IR entity representations with control- and data-flow information~\cite{graph4}. A study models optimization-pass selection as a Markov decision process and enhances program representations by integrating static features with GNN-derived control-flow graph embeddings~\cite{graph5}. PROGRAML presents a unified program-graph representation capturing control and data, and call dependencies using compiler IR, that enables message-passing neural networks to approximate classical compiler analyses~\cite{graph6}. 

In case of the phase ordering problem, researchers introduce a lightweight multi-phase learning framework that builds accurate performance-prediction models from small, diversity-driven samples and then searches the optimization-flag space via an improved particle-swarm algorithm~\cite{phase2}. A related work shows that specialized compiler pass orders can yield significant energy savings on both ARM and x86 multicore systems, independently of improvements in execution time~\cite{phase3}. Through extensive design-space exploration, it reveals that specific phase sequences reduce energy consumption by up to 24\% and that performance-driven orders do not always correspond to energy-optimal ones. AutoPhase applies deep reinforcement learning to the phase-ordering problem in high-level synthesis, using program-state features to guide the sequential application of LLVM optimization passes~\cite{phase4}. The results illustrate that RL achieves up to 16\% circuit-performance improvement over -O3. Researchers propose an RL framework that simultaneously optimizes code size and execution time by modeling programs with IR2Vec embeddings and learning effective pass sequences through Deep Q-Networks~\cite{phase5}. By introducing Oz-based subsequences and the Oz-Dependence Graph, the method outperforms LLVM’s size-oriented optimization levels across SPEC and MiBench benchmarks. MiCOMP clusters LLVM’s O3 passes into sub-sequences and uses machine-learning models to predict performance for complete ordered sequences, which enable efficient exploration of the enormous phase-ordering space~\cite{phase6}. The framework achieves up to 1.51× speedups and obtains 90\% of attainable gains while exploring less than 0.001\% of the optimization space. YaCoS provides a comprehensive compiler-optimization exploration infrastructure with benchmark suites, feature extractors, distance metrics and search algorithms designed to support learning-based phase-ordering methods~\cite{phase7}. The authors demonstrate that machine-learned sequences can outperform clang -Oz by an average of 3.75\% in code size reduction. Another related work introduces booster pass chains, sequences of passes exhibiting positive interplay, and proposes an iterative generate-and-select framework that automatically identifies such chains to construct high-quality candidate pass sequences~\cite{phase1}. Evaluated on 6,186 programs across diverse datasets, the method achieves improvements over LLVM -Oz.

\subsection{Comparison}

Table~\ref{tab:summary} compares MileStone with a wide range of compiler phase-ordering frameworks based on their compiler support, program representation, optimization scope, search-space limitations, multi-objective capability and underlying optimization techniques. Most prior work focuses on a single objective, relies on sequence encodings or IR2Vec embeddings and often restricts the search space to predefined optimization levels such as O3 or Oz. Several approaches apply heuristic search or classic reinforcement learning but lack structural program representations or multi-objective support. MileStone differs from these methods by combining graph-based intermediate representations with an RL engine that uses either DQN or PPO and a self-evolving database for gathering diverse optimization examples. It also supports true multi-objective optimization without limiting the search space and operates on both LLVM and GCC infrastructures. These features make MileStone more flexible, scalable and capable of finding balanced solutions across execution time, code size and energy consumption compared to related methods listed in the table.

\begin{table}[ht]
\caption{Summary of Compiler Phase-Ordering Optimization Approaches}
\label{tab:summary}
\large
\begin{adjustbox}{width=\textwidth}
\begin{tabular}{ccccccccc}
\toprule
\textbf{Paper} & \textbf{\shortstack{Supported \\ Compilers}} & \textbf{\shortstack{Program \\Representation}} & \textbf{\shortstack{Optimization \\Level}} & \textbf{\shortstack{Self-Limited \\Search Space}} & \textbf{\shortstack{Multi-objective \\ Optimization}} & \textbf{\shortstack{Heuristic \\Search Approach}} & \textbf{\shortstack{RL \\Approach}} & \textbf{\shortstack{ML/DL \\Approach}} \\
\midrule
MileStone & GCC, LLVM & CDFG & O3 & NO & YES & - & PPO/DQN & GCN \\
MiCOMP \cite{phase6} & LLVM & Sequence Encoding & O3 & YES & NO & - & - & Supervised Learning \\
POSET-RL \cite{phase5} & LLVM & IR2vec Encoding & Oz & YES & NO & - & DDQN & - \\
FlexPO \cite{phase-prob} & LLVM & CFG & O1, O2, O3, Os, Oz & YES & NO & - & Classic RL & - \\
Shackleton \cite{phase-o} & LLVM & - & O1, O2, O3 & NO & NO & Linear Genetic Programming & - & - \\
CompTuner \cite{phase2} & GCC, LLVM & - & Custom passes & NO & NO & Particle Swarm Optimization & - & Random Forest \\
BPC \cite{phase1} & LLVM & Booster Pass Chains & Oz & YES & NO & - & - & MLP \\
CORL \cite{comp4} & LLVM & Inst2vec Encoding & O3 & NO & NO & - & DQN & - \\
COBAYN \cite{comp3} & GCC & Microarch. Features & O3 & NO & NO & - & - & Bayesian Networks \\
YACOS \cite{phase7} & LLVM & Dynamic Features & Oz & NO & NO & - & - & Supervised Learning \\
\shortstack{Heuristic \\Optimization}~\cite{phase8} & VPO & - & Custom passes & YES & NO & Various (including GA) & - & - \\
\bottomrule
\end{tabular}
\end{adjustbox}
\end{table}

\section{Conclusion}\label{sec:con}

This paper introduced MileStone, a multi-objective framework for compiler phase ordering that combines graph-based program representations, static performance prediction and reinforcement-learning-based exploration. The framework models the phase ordering problem as a trade-off among execution time, code size and energy consumption, and it supports user-defined constraints to guide optimization decisions. MileStone integrates a graph neural network predictor with an adaptive reinforcement-learning engine and a self-evolving database, which together enable fast exploration of large optimization spaces without relying on repeated program execution. Experimental results on standard benchmarks show that MileStone discovers high-quality Pareto-optimal solutions and outperforms LLVM optimization levels, heuristic search methods and standalone reinforcement-learning agents. The framework also meets energy constraints with high accuracy and achieves significant reduction in execution time under fixed energy budgets. These findings demonstrate that combining static graph learning with reinforcement-based search provides an effective and scalable solution for multi-objective compiler optimization. 

As for future research we have a plan to extend MileStone in several directions. One direction is to apply the framework to larger and more diverse program sets, including real-world industrial workloads. Another direction is to incorporate more detailed hardware and microarchitectural models to improve prediction accuracy for different platforms. MileStone can also be enhanced to support dynamic optimization scenarios. Finally, integrating additional learning signals, such as runtime profiling or transfer learning across domains, may further increase the adaptability and robustness of the framework.

\balance

\bibliographystyle{ACM-Reference-Format}
\bibliography{sample-base}

\end{document}